\documentclass[fleqn,usenatbib]{mnras}
\usepackage[T1]{fontenc}
\usepackage{amsmath}
\usepackage{newtxtext,newtxmath} 
\usepackage{bm}
\usepackage{graphicx}
\usepackage{booktabs}

\newcommand{\Msun}{\mathrm{M}_\odot}
\newcommand{\kpc}{\,\mathrm{kpc}}
\newcommand{\rhosun}{\rho_\odot}
\newcommand{\ahat}{\hat{\bm a}}
\newcommand{\bhat}{\hat{\bm b}}
\newcommand{\chat}{\hat{\bm c}}
\newcommand{\nhat}{\hat{\bm n}}
\newcommand{\Rg}{R_{g}}

\title[Triaxiality of the Milky Way halo and $\tau_{\rm LMC}$]
      {Triaxiality of the Milky Way dark matter halo and the microlensing
       optical depth towards the Large Magellanic Cloud}

\author[J. Garc\'ia-Bellido and M. R. S. Hawkins]
{Juan Garc\'ia-Bellido,$^{1}$\thanks{E-mail: juan.garciabellido@uam.es}
 and M. R. S. Hawkins$^{2}$\thanks{E-mail: mrsh@roe.ac.uk}
\\
$^{1}$Instituto de F\'isica Te\'orica UAM/CSIC, Universidad Aut\'onoma de
Madrid, Cantoblanco, 28049 Madrid, Spain\\
$^{2}$Institute for Astronomy, University of Edinburgh, Royal Observatory,
Blackford Hill, Edinburgh EH9 3HJ, UK}

\date{Accepted XXX. Received YYY; in original form ZZZ}

\pubyear{2026}

\begin{document}
\label{firstpage}
\maketitle

\begin{abstract}
The microlensing optical depth towards the Large Magellanic Cloud (LMC),
$\tau_{\rm LMC}$, bounds the compact-object fraction of the Galactic dark
matter halo, but is almost always evaluated for a spherical halo. We compute
it for triaxial haloes and show that the shape dependence is carried entirely
by the projection of the line of sight onto the halo principal axes: for a
scale-free halo $\rho\propto R^{-\alpha}$ at fixed local density it factorises
exactly as
$\tau/\tau_{\rm sph}=S_\odot^{\alpha}\langle S^{-\alpha}\rangle_w$, with
$S(\Omega)$ the ellipsoidal shape function. The LMC lies only $0\fdg78$ from
perpendicular to the Sun-Galactic Centre line, making it a near-optimal probe
of in-plane elongation and a poor probe of vertical flattening. For the halo
shapes favoured by \textit{Gaia} we find
$\tau_{\rm LMC}/\tau_{\rm sph}=0.95$--$1.15$ at fixed local density,
$1.57$ for the Law-Majewski logarithmic potential, and $0.67$--$1.54$ when the
major-axis azimuth is left free. For the Einasto halo required by the declining
\textit{Gaia} DR3 rotation curve, truncation near $25\kpc$ is offset by the
larger local density it implies and $\tau$ falls by only $7$--$15$ per cent;
but the median lens distance moves inwards from $13.4$ to $8.5\kpc$, where the
logarithmic slope is steep and rising, amplifying the shape sensitivity: the
\textit{Gaia} RR~Lyrae enhancement grows from $15$ to $23$ per cent and the
orientation-free range widens to $0.62$--$1.62$. The systematic on the inferred
MACHO or primordial-black-hole fraction is two-sided, of order $\pm15$ per cent
for the measured shapes and a factor $2.6$ under full orientation freedom,
exceeding the statistical errors of the EROS-2 and OGLE bounds.
\end{abstract}


\section{Introduction}
\label{sec:intro}

Microlensing of Magellanic Cloud stars remains the sharpest probe of compact
dark matter in the Galactic halo over the mass range
$10^{-7}\lesssim M/\Msun\lesssim 30$
\citep{Paczynski1986,Griest1991,Alcock2000,Tisserand2007,Wyrzykowski2011a,
Wyrzykowski2011b}. The quantity confronted with observation is the optical
depth
\begin{equation}
\tau = \frac{4\pi G}{c^{2}}\int_{0}^{D_s}\! \mathrm{d}D\,\rho\big(\bm r(D)\big)\,
       \frac{D\,(D_s-D)}{D_s}\,,\qquad
\bm r(D)=\bm r_\odot + D\,\nhat\,,
\label{eq:tau}
\end{equation}
which for an all-MACHO halo is quoted as $\tau_{\rm sph}\simeq4.7\times10^{-7}$
for the standard spherical, cored isothermal `model~S'. The measured value,
$\tau_{\rm obs}=1.2^{+0.4}_{-0.3}\times10^{-7}$ from MACHO \citep{Alcock2000},
and the far more restrictive EROS-2 and OGLE non-detections
\citep{Tisserand2007,Wyrzykowski2011a,Wyrzykowski2011b}, are translated into a
halo fraction $f=\tau_{\rm obs}/\tau_{\rm sph}$. Every such bound, including
the widely used exclusion of stellar-mass primordial black holes (PBHs) as the
dominant dark matter component \citep{Carr2023,GreenKavanagh2021}, is therefore
only as good as the halo model in the integrand
\citep{Hawkins2015,Calcino2018,GarciaBellido2024}.

That halo model is almost universally taken to be spherical. The Milky Way
halo, however, is not. \textit{Gaia} astrometry of RR~Lyrae shows the inner
stellar halo to be triaxial with its longest axis lying in the disc plane,
misaligned by $\sim70^\circ$ with the Sun--Galactic Centre direction
\citep{Iorio2018,IorioBelokurov2019}; the stellar halo is additionally tilted
out of the plane \citep{Han2022}, and equilibrium orbit-superposition models of
the \textit{Gaia} Sausage--Enceladus (GSE) debris require a near-prolate dark
halo with short-to-long axis ratio $q_{\rm dm}=0.87^{+0.05}_{-0.09}$ whose long
axis is inclined at $\beta_{\rm dm}=43^{+22}_{-8}$ degrees to the disc
\citep{Dillamore2026}. Stream modelling has long pointed the same way
\citep{LawMajewski2010,Vera-Ciro2013,Vasiliev2021}, and the innermost halo has
now been measured to be mildly triaxial as well \citep{Woudenberg2024}.

Halo flattening was recognised early as a source of systematic error in $\tau$
\citep{SackettGould1993,Frieman1994}, but those analyses predate any
measurement of the halo's orientation, and treated axisymmetric (oblate or
prolate about $\hat{\bm z}$) configurations only. The first point of this paper
is that the orientation, not merely the axis ratios, is what matters, because
the LMC line of sight happens to lie almost exactly perpendicular to the
Sun--Galactic Centre axis. We quantify the effect and give the geometry
explicitly.

A second, independent revision of the halo model has arrived from the same data
set. \textit{Gaia} DR3 kinematics of disc tracers out to $\sim28\kpc$ reveal a
Keplerian decline of the circular velocity curve beyond $\sim19\kpc$, implying
a total Galactic mass of only $2.06^{+0.24}_{-0.13}\times10^{11}\Msun$
\citep{Jiao2023}, with an independent analysis favouring a cored Einasto halo
of virial mass $1.81^{+0.06}_{-0.05}\times10^{11}\Msun$ \citep{Ou2024}; the
robustness of the inference is debated \citep{Koop2024}. This is precisely the
halo model recently used to relax the MACHO and PBH bounds
\citep{GarciaBellido2024}. Our second point is that the two revisions are not
independent corrections to be applied in sequence. A truncated halo is not
simply a fainter lens: it moves the microlensing weight inwards, into radii
where the logarithmic density slope is steep and rapidly varying, and since the
shape response is governed by that slope, it makes $\tau_{\rm LMC}$
\emph{more}, not less, sensitive to triaxiality. We find the amplification to
be a factor $\simeq1.5$, so that the shape systematic and the mass-model
revision reinforce one another.

We stress at the outset that the correction we compute is two-sided. Depending
on the orientation of the halo major axis relative to the LMC azimuth,
triaxiality can either raise or lower $\tau_{\rm LMC}$, and therefore either
weaken or strengthen the resulting exclusion. It happens to weaken it for the
orientation currently favoured by \textit{Gaia}, but that is a property of the
measured geometry and not of triaxiality as such.

The paper is organised as follows. Section~\ref{sec:geom} fixes the geometry
and tabulates the projections of the Earth--LMC line of sight onto the halo
principal axes; Section~\ref{sec:models} defines the halo families and the
normalisation and ellipsoidal-radius conventions; Section~\ref{sec:factor}
derives the exact scale-free factorisation; Section~\ref{sec:results} gives the
flat-rotation-curve results and Section~\ref{sec:decline} those for the
declining \textit{Gaia} DR3 rotation curve. Section~\ref{sec:caveats} collects
the assumptions and their consequences, and Section~\ref{sec:concl} the
implications for the MACHO and PBH bounds.

\section{Geometry and projections onto the halo axes}
\label{sec:geom}

We work in a right-handed Galactocentric Cartesian frame $(X,Y,Z)$ with
$\hat{\bm X}$ pointing from the Sun towards the Galactic Centre, $\hat{\bm Y}$
towards $\ell=90^\circ$ and $\hat{\bm Z}$ towards the North Galactic Pole, so
that the Sun sits at $\bm r_\odot=(-R_0,0,z_\odot)$. We adopt
$R_0=8.122\pm0.031\kpc$, the value returned by the S2 orbit fit of
\citet{Gravity2018}, together with $z_\odot=20.8\pm0.3\,$pc
\citep{BennettBovy2019}. The dedicated geometric determination of
\citet{Gravity2019}, $R_0=8178\pm13\,({\rm stat})\pm22\,({\rm sys})\,$pc, is
consistent with it, and nothing below is sensitive to the difference: varying
$R_0$ over $8.12$--$8.50\kpc$ moves the Galactocentric azimuth of the LMC by
$0\fdg53$, its Galactocentric distance by $5\,$pc, and every shape ratio quoted
in this paper by less than $0.6$ per cent. The unit vector from the Earth to
the LMC, at $(\ell,b)=(280\fdg4652,-32\fdg8884)$, is
\begin{align}
\nhat&=(\cos b\cos\ell,\ \cos b\sin\ell,\ \sin b) \nonumber \\
     &=(+0.15253,\,-0.82576,\,-0.54300)\,.
\label{eq:nhat}
\end{align}
With $D_{\rm LMC}=49.59\kpc$ \citep{Pietrzynski2019} the Galactocentric
position and direction of the LMC are ($r_{\rm LMC}=49.0\kpc$)
\begin{align}
\bm r_{\rm LMC}&=(-0.558,\,-40.950,\,-26.907)\kpc, \nonumber\\
\hat{\bm r}_{\rm LMC}&=(-0.01139,\,-0.83568,\,-0.54910)\,.
\label{eq:rlmc}
\end{align}
Two features of equations~(\ref{eq:nhat})--(\ref{eq:rlmc}) drive everything
that follows. First, the LMC lies at Galactocentric azimuth
$\phi_{\rm LMC}=\arctan(Y/X)=269\fdg22$, i.e.\ only $0\fdg78$ from the
$-\hat{\bm Y}$ axis: the line of sight is essentially \emph{perpendicular} to
the Sun--Galactic Centre direction, so the halo density it samples is that
along a direction $89\fdg2$ away (modulo $180^\circ$) from the direction in
which the local density $\rhosun$ is measured. Second, the heliocentric and
Galactocentric directions differ by only $9\fdg43$, and $r_{\rm LMC}\gg R_0$,
so the line of sight is close to radial from the Galactic Centre over most of
its length; the residual difference is nevertheless responsible for a genuine
asymmetry between mirror-image halo orientations
(Section~\ref{sec:results}).

A triaxial halo is specified by its principal axes $(\ahat,\bhat,\chat)$,
major, intermediate and minor, and by the axis ratios $p\equiv b/a$ and
$q\equiv c/a$, so that the density is a function of the ellipsoidal radius
\begin{equation}
R^{2}=(\bm r\!\cdot\!\ahat)^{2}
     +(\bm r\!\cdot\!\bhat)^{2}/p^{2}
     +(\bm r\!\cdot\!\chat)^{2}/q^{2}\,,
\label{eq:R}
\end{equation}
normalised so that $R$ equals the distance along the major axis. We
parametrise the orientation by the azimuth $\gamma$ of $\ahat$ measured from
$\hat{\bm X}$ towards $\hat{\bm Y}$, its elevation $\beta$ above the disc
plane, and a roll $\psi$ of $\bhat$ about $\ahat$. Because the ellipsoid is
invariant under $\bm r\to-\bm r$, $\gamma$ has period $180^\circ$.

The projections required by equation~(\ref{eq:R}) are the direction cosines
$\nhat\!\cdot\!\ahat$, $\nhat\!\cdot\!\bhat$, $\nhat\!\cdot\!\chat$ and their
Galactocentric counterparts
$\hat{\bm r}_{\rm LMC}\!\cdot\!(\ahat,\bhat,\chat)$; they are collected in
Table~\ref{tab:proj} for a set of measured halo shapes. It is convenient to
define the dimensionless shape function
\begin{equation}
S(\Omega)\equiv\frac{R}{r}
=\Big[(\hat{\bm r}\!\cdot\!\ahat)^{2}
     +(\hat{\bm r}\!\cdot\!\bhat)^{2}/p^{2}
     +(\hat{\bm r}\!\cdot\!\chat)^{2}/q^{2}\Big]^{1/2},
\label{eq:S}
\end{equation}
so that $S_\odot\equiv S(\Omega_\odot)$ and
$S_{\rm LMC}\equiv S(\Omega_{\rm LMC})$ measure how far the Sun and the LMC
sit, in units of the major-axis radius, from the isodensity surface passing
through them.

\begin{table*}
\caption{Projection of the Earth$\to$LMC line of sight $\nhat$ and of the
Galactocentric LMC direction $\hat{\bm r}_{\rm LMC}$ onto the principal axes
$(\ahat,\bhat,\chat)$ of the Milky Way dark matter halo, for representative
measured halo shapes. Angles (deg) are with respect to the axis line
(i.e.\ $\arccos|\!\cos\!|$). $S_\odot$, $S_{\rm LMC}$ are the shape functions
of equation~(\ref{eq:S}), evaluated at the Sun and along the LMC direction.
Orientation angles $(\gamma,\beta,\psi)$ in degrees. The last column gives the
radial range over which each shape is actually constrained by its source; the
LMC line of sight extends to $49.6\kpc$, so all entries involve some
extrapolation (Section~\ref{sec:caveats}).}
\label{tab:proj}
\centering
\footnotesize
\setlength{\tabcolsep}{2.5pt}
\begin{tabular}{lcccccccc}
\hline
Halo shape & $(p,q)$ & $(\gamma,\beta,\psi)$ &
$\nhat\!\cdot\!(\ahat,\bhat,\chat)$ & angles [deg] &
$\hat{\bm r}_{\rm LMC}\!\cdot\!(\ahat,\bhat,\chat)$ &
$(S_\odot,S_{\rm LMC})$ & valid $r$ [kpc] \\
\hline
Spherical
 & $(1.000,1.000)$ & $(0,0,0)$
 & $(+0.153,-0.826,-0.543)$ & $(81.2,34.3,57.1)$
 & $(-0.011,-0.836,-0.549)$ & $(1.000,1.000)$ & ---\\
Oblate, $z$ minor
 & $(1.000,0.800)$ & $(0,0,0)$
 & $(+0.153,-0.826,-0.543)$ & $(81.2,34.3,57.1)$
 & $(-0.011,-0.836,-0.549)$ & $(1.000,1.082)$ & illustrative\\
Oblate, $z$ minor
 & $(1.000,0.600)$ & $(0,0,0)$
 & $(+0.153,-0.826,-0.543)$ & $(81.2,34.3,57.1)$
 & $(-0.011,-0.836,-0.549)$ & $(1.000,1.239)$ & illustrative\\
Prolate-$z^{\dagger}$
 & $(0.841,0.831)$ & $(0,90,0)$
 & $(-0.543,-0.826,-0.153)$ & $(57.1,34.3,81.2)$
 & $(-0.549,-0.836,+0.011)$ & $(1.204,1.135)$ & $\lesssim15$--$20$\\
IB19
 & $(0.787,0.700)$ & $(70,0,0)$
 & $(-0.724,-0.426,-0.543)$ & $(43.6,64.8,57.1)$
 & $(-0.789,-0.275,-0.549)$ & $(1.242,1.166)$ & $\lesssim30$\\
IB19, mirrored
 & $(0.787,0.700)$ & $(110,0,0)$
 & $(-0.828,+0.139,-0.543)$ & $(34.1,82.0,57.1)$
 & $(-0.781,+0.297,-0.549)$ & $(1.242,1.170)$ & $\lesssim30$\\
IB19 $+$ tilt
 & $(0.787,0.700)$ & $(70,0,20)$
 & $(-0.724,-0.586,-0.365)$ & $(43.6,54.1,68.6)$
 & $(-0.789,-0.446,-0.422)$ & $(1.259,1.144)$ & $\lesssim30$\\
GSE prolate$^{\ast}$
 & $(0.870,0.870)$ & $(90,43,0)$
 & $(-0.974,-0.153,+0.166)$ & $(13.0,81.2,80.4)$
 & $(-0.986,+0.011,+0.168)$ & $(1.149,1.005)$ & $6$--$60$\\
GSE prolate, $\gamma=0$
 & $(0.870,0.870)$ & $(0,43,0)$
 & $(-0.259,-0.826,-0.501)$ & $(75.0,34.3,59.9)$
 & $(-0.383,-0.836,-0.394)$ & $(1.073,1.129)$ & $6$--$60$\\
Law--Majewski$^{\ddagger}$
 & (potential) & $(97,0,90)$
 & $(-0.838,-0.543,+0.051)$ & $(33.1,57.1,87.1)$
 & $(-0.828,-0.549,-0.113)$ & --- & $20$--$60$\\
\hline
\end{tabular}
\begin{flushleft}
Sources: Prolate-$z$, \citet{Woudenberg2024}; IB19 and IB19$+$tilt,
\citet{IorioBelokurov2019}, with the out-of-plane tilt of \citet{Han2022};
GSE prolate, \citet{Dillamore2026}; Law--Majewski, \citet{LawMajewski2010}.\\
$^{\ast}$\,The azimuth $\gamma$ of the GSE-prolate halo is \emph{not}
determined by \citet{Dillamore2026}, who assume only that the long axes of the
two haloes and the disc normal are coplanar. The two values bracket the
possibilities; $\gamma=90^\circ$ aims the long axis at the LMC azimuth and is
therefore a maximal, not a measured, configuration.\\
$^{\dagger}$\,\citet{Woudenberg2024} quote $(p,q)=(1.013,1.204)$ in a
convention normalised to the $x$ axis. Reordering their axes by length and
renormalising to the major axis (here $\hat{\bm Z}$, since $q>1$ in their
convention) gives the $(0.841,0.831)$ and $(\gamma,\beta,\psi)=(0,90,0)$ used
here.\\
$^{\ddagger}$\,\citet{LawMajewski2010} assign their minor, intermediate and
major axes to $(\ell,b)=(7,0)$, $(0,90)$ and $(97,0)$ respectively. The entry
above reproduces that assignment: $\psi=90^\circ$ places the intermediate axis
at the North Galactic Pole and the minor axis in the disc plane at
$\ell=7^\circ$.
\end{flushleft}
\end{table*}

Table~\ref{tab:proj} already makes the central geometric point. For the
\textit{Gaia} RR~Lyrae orientation the line of sight is only $43\fdg6$ from the
halo major axis, against $81\fdg2$ for a halo whose major axis lies along the
Sun--Galactic Centre line; for the tilted GSE-prolate halo with its long axis
pointing at the LMC azimuth the misalignment collapses to $13\fdg0$. Since
$\tau$ weights the density along $\nhat$ while the observational normalisation
$\rhosun$ is imposed along $\hat{\bm r}_\odot$, the ratio
$S_\odot/S_{\rm LMC}$ is the natural expansion parameter, and it deviates from
unity by $5$--$25$ per cent for all measured shapes.

\section{Halo models, normalisation and numerical implementation}
\label{sec:models}

We consider two families with ellipsoidal isodensity surfaces,
\begin{align}
\rho_{\rm iso}(R)&=\frac{\rho_0\,(R_0^{2}+a_c^{2})}{R^{2}+a_c^{2}},
 \qquad a_c=5\kpc, \label{eq:iso}\\
\rho_{\rm NFW}(R)&=\frac{\rho_s}{(R/r_s)\,(1+R/r_s)^{2}},
 \qquad r_s=18\kpc, \label{eq:nfw}
\end{align}
the latter being the profile of \citet{NFW1997}, and, separately, the triaxial
logarithmic potential used in Sagittarius-stream fits
\citep{LawMajewski2010},
\begin{equation}
\Phi=\tfrac{1}{2}v_h^{2}\ln\!\big(u_a^{2}+u_b^{2}/p_\Phi^{2}
        +u_c^{2}/q_\Phi^{2}+r_h^{2}\big),
\label{eq:log}
\end{equation}
whose density we obtain exactly from Poisson's equation,
$\rho=\nabla^{2}\Phi/4\pi G$,
\begin{equation}
\rho=\frac{v_h^{2}}{8\pi G}\!\left[
\frac{2\big(1+p_\Phi^{-2}+q_\Phi^{-2}\big)}{Q}
-\frac{4\big(u_a^{2}+u_b^{2}p_\Phi^{-4}+u_c^{2}q_\Phi^{-4}\big)}{Q^{2}}\right],
\label{eq:logrho}
\end{equation}
with $Q=u_a^{2}+u_b^{2}/p_\Phi^{2}+u_c^{2}/q_\Phi^{2}+r_h^{2}$. For the
spherical case equation~(\ref{eq:iso}) with
$\rho_0=0.0079\,\Msun\,{\rm pc}^{-3}$, $R_0=8.5\kpc$ and $D_s=49.59\kpc$
reproduces the canonical $\tau_{\rm sph}=4.69\times10^{-7}$ to three digits,
validating our quadrature; the value usually quoted for model~S assumes
$D_s=50\kpc$, for which we obtain $4.72\times10^{-7}$. With the modern
$R_0=8.122\kpc$ we obtain $\tau_{\rm sph}=4.50\times10^{-7}$, and the NFW halo
normalised to $\rhosun=0.0093\,\Msun\,{\rm pc}^{-3}$ gives
$4.86\times10^{-7}$. These two local densities are the historical values
attached to the two families; modern determinations cluster somewhat higher,
around $0.010$--$0.013\,\Msun\,{\rm pc}^{-3}$, and since $\tau\propto\rhosun$
and $f\propto1/\tau$ this is a first-order systematic on the absolute optical
depths comparable in size to the shape effect studied here. All the
\emph{ratios} we quote are, by construction, independent of it, and we
therefore work throughout with ratios. For the same reason
$\tau_{\rm iso}$ and $\tau_{\rm NFW}$ in Table~\ref{tab:tau} should not be
compared with one another directly. All optical depths below are for $f=1$
and scale linearly with the compact-object fraction.

To determine how much triaxiality modifies $\tau$ we need to state what is held
fixed. We explore two choices:
\begin{enumerate}
\item[(i)] \emph{fixed local density}: the normalisation is set by
$\rho(\bm r_\odot)=\rhosun$, the observationally robust anchor;
\item[(ii)] \emph{fixed enclosed mass}: the normalisation is set by
$M_{\rm h}(<50\kpc)$ evaluated over a \emph{sphere} of radius $50\kpc$ by
direct three-dimensional quadrature. For the cored isothermal reference model
at $R_0=8.122\kpc$ this is $3.85\times10^{11}\Msun$ (at $R_0=8.5\kpc$ it would
be $4.12\times10^{11}\Msun$); each halo family is normalised to its own
spherical value, not to that of model~S.
\end{enumerate}
These bracket the physically reasonable options and, as we now show, they
respond to triaxiality in opposite ways.

A third convention must be fixed once the halo is not scale-free.
Equation~(\ref{eq:R}) normalises $R$ to the major axis, so that $R\ge r$
everywhere and deforming a sphere at fixed $R$ also shrinks it. The alternative
is the volume-preserving radius $\Rg\equiv(pq)^{1/3}R$, for which the
isodensity ellipsoid of radius $\Rg$ encloses the same volume as a sphere of
that radius. For a scale-free halo the choice is immaterial:
equation~(\ref{eq:factor}) below is invariant under $S\to\lambda S$, and we
have verified numerically that $\tau$ is unchanged to machine precision
($\lesssim10^{-15}$) for $\alpha=2,4,8$. For the cored isothermal halo the two
conventions differ by up to $5$ per cent, but for the sharply truncated Einasto
halo of Section~\ref{sec:decline} they differ by tens of per cent.

\emph{Because this convention dependence is one of the results of this paper,
we state it explicitly at every number we quote.} Sections~\ref{sec:results}
and the first six columns of Table~\ref{tab:tau} use the major-axis radius $R$;
Section~\ref{sec:decline} and Table~\ref{tab:ein} use $\Rg$ throughout, as the
truncated profile requires. The last two columns of Table~\ref{tab:tau} repeat
the flat-rotation-curve ratios in the $\Rg$ convention, so that every number
quoted in Section~\ref{sec:decline} can be traced to a tabulated value in the
same convention.

\subsection{Numerical implementation}

The line-of-sight integral in equation~(\ref{eq:tau}) is evaluated by
Gauss--Legendre quadrature with $400$ nodes; the enclosed mass of
normalisation~(ii) uses a $80\times40\times64$ product rule in
$(r,\cos\theta,\phi)$. Convergence was verified at the $10^{-6}$ level by
doubling the number of nodes in each dimension. The factorisation of
equation~(\ref{eq:factor}) was verified against direct quadrature to machine
precision.

\section{An exact factorisation}
\label{sec:factor}

For a scale-free halo, $\rho(R)\propto R^{-\alpha}$, normalisation~(i) gives
$\rho(\bm r)=\rhosun\,[R_\odot/R(\bm r)]^{\alpha}
 =\rhosun\,[R_0S_\odot/(r\,S(\Omega))]^{\alpha}$,
and equation~(\ref{eq:tau}) factorises exactly:
\begin{equation}
\frac{\tau}{\tau_{\rm sph}}
= S_\odot^{\alpha}\,\big\langle S(\Omega)^{-\alpha}\big\rangle_{w}, \qquad
w(D)=\frac{D(D_s-D)}{r(D)^{\alpha}D_s}\,,
\label{eq:factor}
\end{equation}
where $\langle\cdot\rangle_w$ is the $w$-weighted average along the line of
sight. All halo-shape dependence is thus carried by the shape function $S$
evaluated at the Sun and along $\nhat$, that is, precisely by the projections
of Table~\ref{tab:proj}. We have verified equation~(\ref{eq:factor})
numerically to machine precision for $\alpha=2$. The crude single-direction
estimate $\tau/\tau_{\rm sph}\approx(S_\odot/S_{\rm LMC})^{\alpha}$ reproduces
the exact result to $5$--$10$ per cent for mild triaxiality but fails for
strongly elongated haloes, because the line of sight sweeps Galactocentric
azimuths from $0^\circ$ (at the Sun) to $89\fdg2$ (at the LMC) and therefore
does not sample a single direction $\Omega$.

That azimuthal sweep breaks the mirror symmetry
$\gamma\to180^\circ-\gamma$: a major axis at $\gamma=70^\circ$ is crossed by
the line of sight, whereas at $\gamma=110^\circ$ it is approached only at the
far end. The two orientations are equally distant from the LMC azimuth
($19\fdg2$ and $20\fdg8$) and have identical $S_\odot/S_{\rm LMC}$ to three
digits (Table~\ref{tab:proj}), yet give $\tau/\tau_{\rm sph}=1.100$ and
$0.962$ respectively, a $14$ per cent difference that no direction-cosine
argument alone can capture. Since the sign of $\gamma$ is a frame convention
that is not always stated unambiguously in the literature, this is a real and
easily overlooked ambiguity.

\section{Results for a flat rotation curve}
\label{sec:results}

\begin{figure}
\includegraphics[width=\columnwidth]{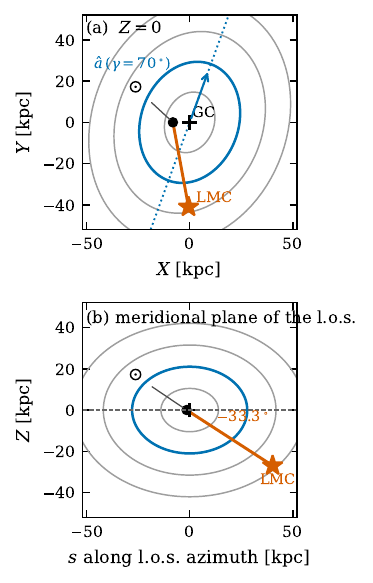}
\caption{Geometry of the LMC line of sight (orange) through a triaxial halo
with the \textit{Gaia} RR~Lyrae shape $p=0.787$, $q=0.700$,
$\gamma=70^\circ$. Grey curves are isodensity contours at
$R=15,30,45,60\kpc$; the $R=30\kpc$ contour is highlighted. (a) Galactic
plane. (b) The meridional plane containing the line of sight and
$\hat{\bm Z}$. The line of sight is nearly perpendicular to the Sun--Galactic
Centre direction and dives $33\fdg3$ below the disc.}
\label{fig:geom}
\end{figure}

\begin{figure}
\includegraphics[width=\columnwidth]{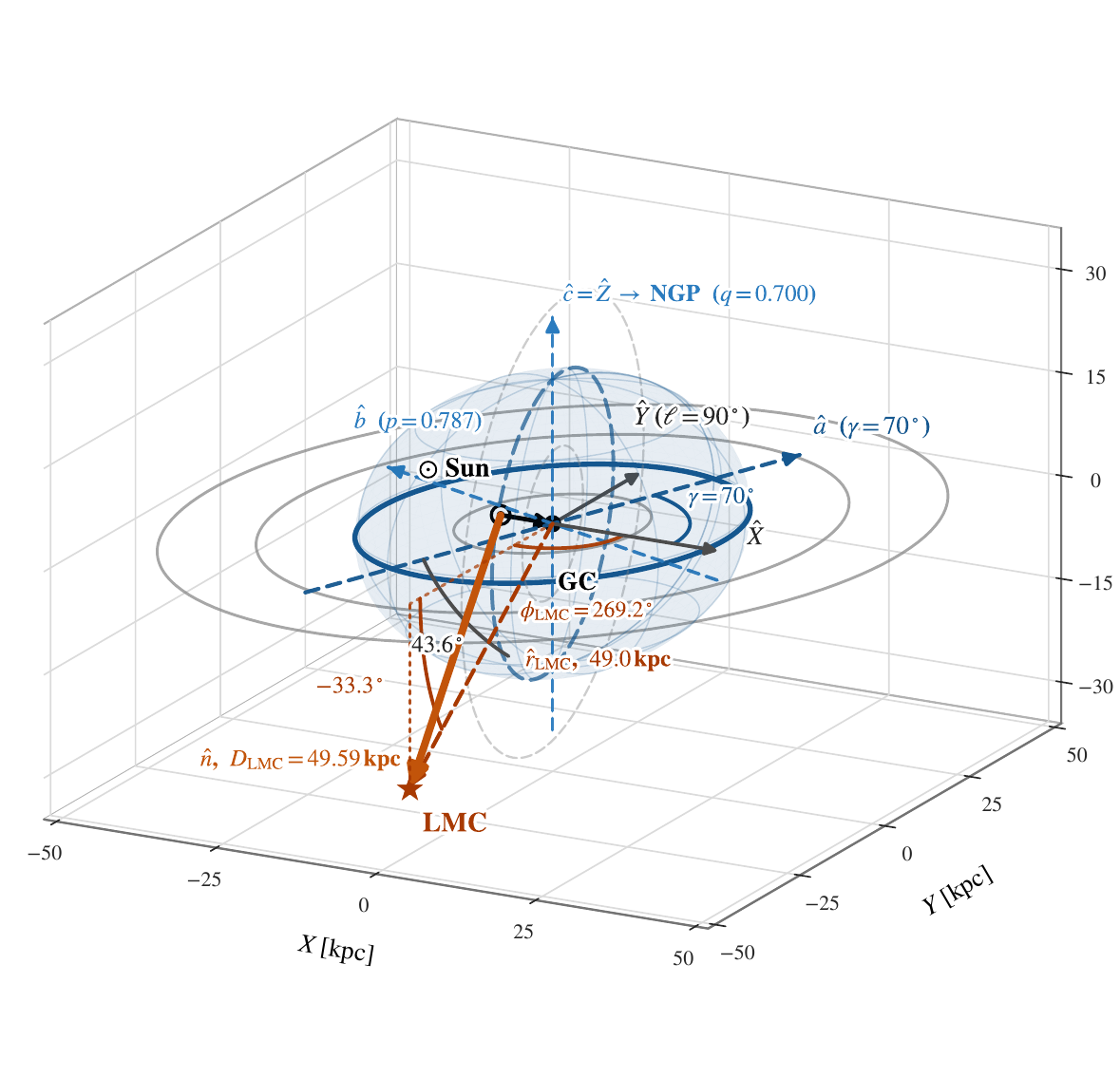}
\caption{Three-dimensional view of the geometry of Fig.~\ref{fig:geom}, in the
right-handed Galactocentric frame $(X,Y,Z)$ of Section~\ref{sec:geom}, for the
\textit{Gaia} RR~Lyrae shape $p=0.787$, $q=0.700$, $\gamma=70^\circ$
\citep{IorioBelokurov2019}. Grey curves are ellipsoidal isodensity contours of
equation~(\ref{eq:R}) in the Galactic plane (solid) and in the meridional plane
containing the line of sight and $\hat{Z}$ (dashed); the $R=30\kpc$ contours
are highlighted in blue and the corresponding isodensity ellipsoid is shown
translucent. Blue dashed lines are the halo principal axes. The thick orange
arrow is the Earth$\rightarrow$LMC line of sight $\nhat$, the dashed orange
arrow the Galactocentric direction $\hat{\bm r}_{\rm LMC}$. Numerical values
for this configuration are given in Tables~\ref{tab:proj} and~\ref{tab:tau}.}
\label{fig:geometry3d}
\end{figure}

\begin{figure*}
\includegraphics[width=\textwidth]{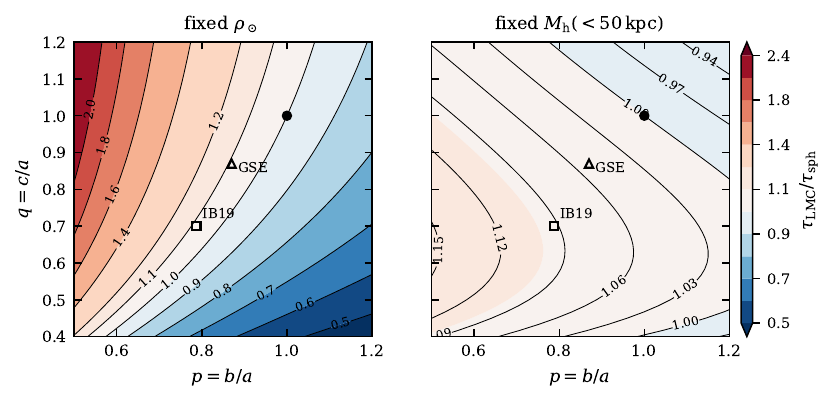}
\caption{Optical depth relative to the spherical halo as a function of the axis
ratios $p=b/a$ and $q=c/a$, for a cored isothermal halo whose major axis is
aimed at the LMC azimuth ($\gamma=89\fdg2$, $\beta=\psi=0$), in the major-axis
radius convention of equation~(\ref{eq:R}). Left: local density $\rhosun$ held
fixed. Right: halo mass within $50\kpc$ held fixed. Note the compressed contour
range on the right. Open symbols mark the \textit{Gaia} RR~Lyrae shape
\citep[IB19;][]{IorioBelokurov2019} and the GSE prolate halo
\citep{Dillamore2026}; the filled circle is the spherical reference.}
\label{fig:pq}
\end{figure*}

\begin{figure}
\includegraphics[width=\columnwidth]{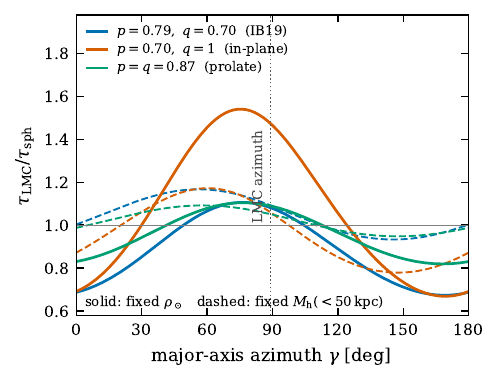}
\caption{Dependence on the azimuth $\gamma$ of the halo major axis, for three
shapes, in the major-axis radius convention. Solid: fixed $\rhosun$. Dashed:
fixed $M_{\rm h}(<50\kpc)$. The dotted line marks the Galactocentric azimuth of
the LMC, $89\fdg2$; the maximum of $\tau$ falls at $\gamma\simeq76^\circ$, not
at the LMC azimuth, because of the azimuthal sweep of the line of sight
discussed in Section~\ref{sec:factor}.}
\label{fig:gamma}
\end{figure}

Table~\ref{tab:tau} collects the optical depths. All numbers in this section
are in the major-axis convention of equation~(\ref{eq:R}) unless stated
otherwise. At fixed $\rhosun$ the measured halo shapes give
\begin{equation}
0.95\ \le\ \tau_{\rm LMC}/\tau_{\rm sph}\ \le\ 1.15\,,
\end{equation}
with the largest enhancement for the GSE-prolate halo whose long axis is tilted
$43^\circ$ out of the plane and is here \emph{assumed} to point towards the LMC
azimuth ($\nhat\!\cdot\!\ahat=-0.974$, i.e.\ $13^\circ$ misalignment); the same
shape with $\gamma=0$ gives $0.953$, so the upper end of this range reflects an
orientation that is not determined by the data (Section~\ref{sec:caveats}). At
fixed $M_{\rm h}(<50\kpc)$ the same shapes give $0.97$--$1.15$. In the
volume-preserving convention the two ranges become $0.98$--$1.18$ and
$0.96$--$1.13$ respectively (last two columns of Table~\ref{tab:tau}). The
isothermal and NFW families agree to better than $1.5$ per cent in the ratio
for every measured shape (the strongly oblate $q=0.6$ illustration differs by
$2.0$ per cent), confirming that the effect is geometric and not
profile-specific. Allowing $\gamma$ to run over its full range at fixed axis
ratios (Fig.~\ref{fig:gamma}) opens
$\tau/\tau_{\rm sph}\in[0.67,1.11]$ for the \textit{Gaia} shape and
$[0.67,1.54]$ for a purely in-plane flattening $p=0.70$, $q=1$; in the
volume-preserving convention these become $[0.73,1.16]$ and $[0.70,1.57]$.

The Law--Majewski triaxial logarithmic potential is the most extreme case. Its
major axis points to $(\ell,b)=(97^\circ,0)$, only $8^\circ$ from the LMC
azimuth, and its potential flattening $(c/a)_\Phi=0.72$ corresponds through
Poisson's equation to an extraordinarily flattened \emph{density}: we find
$\rho_c/\rho_a=0.25$, $0.097$ and $0.053$ at $r=15$, $30$ and $50\kpc$,
approaching the positivity limit $q_\Phi>1/\sqrt2$. We have checked that
equation~(\ref{eq:logrho}) nevertheless remains positive everywhere along the
LMC line of sight out to $60\kpc$, with a minimum of
$5.3\times10^{-4}\,\Msun\,{\rm pc}^{-3}$ at the far end. Normalised to the same
$\rhosun$ this yields $\tau_{\rm LMC}=7.90\times10^{-7}$, a factor $1.573$
above the \emph{spherical logarithmic} halo normalised in the same way
($5.02\times10^{-7}$); note that this reference differs from the cored
isothermal $\tau_{\rm sph}$ used elsewhere in Table~\ref{tab:tau}.
Interchanging major and minor axes gives $0.574$ instead. We regard $1.57$ as
an upper envelope rather than a preferred value: it illustrates that
potential-level flattenings cannot be transcribed into density flattenings
without care, and it is obtained by extrapolating a model fitted over
$20$--$60$ \citep{LawMajewski2010} into the inner halo, where $82$ per cent
of $\tau$ is accumulated (Section~\ref{sec:caveats}).

Linearising about sphericity at fixed $\rhosun$, with the major axis aimed at
the LMC, we obtain the response coefficients
\begin{align}
\partial\ln\tau/\partial\ln p&=-1.02\ ({\rm iso}),\quad -1.13\ ({\rm NFW}),
\nonumber\\
\partial\ln\tau/\partial\ln q&=+0.40\ ({\rm iso}),\quad +0.42\ ({\rm NFW}),
\label{eq:coeffs}
\end{align}
in the major-axis convention, and $-1.12$, $-1.23$, $+0.29$, $+0.33$ in the
volume-preserving convention; the latter are the values quoted for comparison
in Section~\ref{sec:decline}. With the major axis along the Sun--Galactic
Centre line instead, $\partial\ln\tau/\partial\ln p=+0.92$. At fixed
$M_{\rm h}(<50\kpc)$ the same coefficients drop in magnitude to $-0.19$ (iso)
and $-0.21$ (NFW): the enclosed-mass constraint cancels most of the shape
dependence, because elongating the halo towards the LMC while conserving mass
removes density elsewhere. This is the quantitative statement of the degeneracy
between halo shape and halo normalisation, and it is why the two panels of
Fig.~\ref{fig:pq} look so different. Over the range $0.5\le p,q\le1.5$ plotted
there, $\tau/\tau_{\rm sph}$ spans $0.44$--$2.43$ at fixed $\rhosun$ but only
$0.81$--$1.15$ at fixed mass.

\begin{table*}
\caption{Microlensing optical depth towards the LMC for $f=1$, flat rotation
curve. $\bar R$ denotes $\tau/\tau_{\rm sph}$; subscripts $\rho$ and $M$ refer
to normalisation at fixed local density and at fixed $M_{\rm h}(<50\kpc)$
respectively. Columns 2--6 use the major-axis ellipsoidal radius $R$ of
equation~(\ref{eq:R}); the last two columns repeat the isothermal ratios in the
volume-preserving convention $\Rg$, for direct comparison with
Table~\ref{tab:ein}. $\tau_{\rm iso}$ and $\tau_{\rm NFW}$ are normalised to
different local densities (Section~\ref{sec:models}) and should not be compared
directly; the ratios are independent of that choice. The Law--Majewski row is a
potential, not a density ellipsoid, so the $\Rg$ convention does not apply to
it and its ratio is taken with respect to the spherical logarithmic halo
($5.02\times10^{-7}$) rather than to $\tau_{\rm sph}$. The last column is the
MACHO fraction implied by $\tau_{\rm obs}=1.2\times10^{-7}$ \citep{Alcock2000}.}
\label{tab:tau}
\centering
\small
\setlength{\tabcolsep}{2.8pt}
\begin{tabular}{lccccccccc}
\hline
Halo shape & $\tau_{\rm iso}$ & $\bar R^{\rm iso}_{\rho}$ &
$\bar R^{\rm iso}_{M}$ & $\tau_{\rm NFW}$ & $\bar R^{\rm NFW}_{\rho}$ & $f$ &
$\bar R^{\rm iso}_{\rho}(\Rg)$ & $\bar R^{\rm iso}_{M}(\Rg)$ \\
 & $[10^{-7}]$ & & & $[10^{-7}]$ & & & & \\
\hline
Spherical                     & 4.50 & 1.000 & 1.000 & 4.86 & 1.000 & 0.267 & 1.000 & 1.000\\
Oblate $q=0.8$                & 4.06 & 0.902 & 1.039 & 4.35 & 0.895 & 0.296 & 0.924 & 1.032\\
Oblate $q=0.6$                & 3.38 & 0.750 & 1.055 & 3.57 & 0.735 & 0.356 & 0.799 & 1.038\\
Prolate-$z$                   & 4.63 & 1.028 & 0.971 & 5.00 & 1.030 & 0.259 & 1.059 & 0.964\\
IB19 $\gamma=70^\circ$        & 4.95 & 1.100 & 1.153 & 5.39 & 1.110 & 0.242 & 1.152 & 1.134\\
IB19 $\gamma=110^\circ$       & 4.33 & 0.962 & 1.008 & 4.64 & 0.956 & 0.277 & 1.015 & 0.998\\
IB19 $+$ tilt $\psi=20^\circ$ & 5.04 & 1.120 & 1.147 & 5.49 & 1.130 & 0.238 & 1.172 & 1.130\\
GSE $\beta=43^\circ$, $\gamma=90^\circ$ & 5.17 & 1.149 & 1.108 & 5.64 & 1.162 & 0.232 & 1.176 & 1.100\\
GSE $\beta=43^\circ$, $\gamma=0^\circ$  & 4.29 & 0.953 & 1.023 & 4.63 & 0.953 & 0.280 & 0.979 & 1.014\\
Law--Majewski                 & 7.90 & 1.573 & 0.945 & ---  & ---   & 0.152 & ---   & ---\\
\hline
\end{tabular}
\end{table*}

\section{The declining \textit{Gaia} DR3 rotation curve}
\label{sec:decline}

\begin{figure*}
\includegraphics[width=\textwidth]{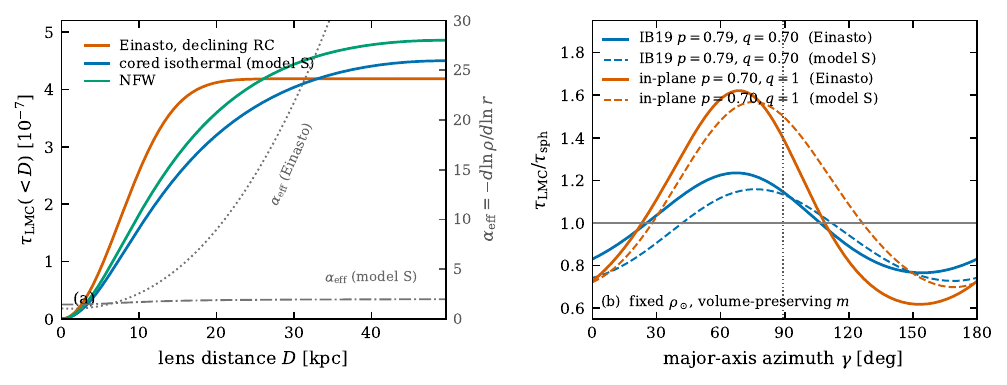}
\caption{(a) Cumulative optical depth $\tau_{\rm LMC}(<D)$ for the Einasto halo
demanded by the declining \textit{Gaia} DR3 rotation curve
\citep{Jiao2023,GarciaBellido2024}, compared with the cored isothermal model~S
and an NFW halo, each normalised to its own local density. Grey curves (right
axis) show the effective logarithmic slope $\alpha_{\rm eff}(r)$ along the line
of sight. The Einasto integral saturates by $D\simeq25\kpc$, and half of it is
accumulated inside $8.5\kpc$. (b) Orientation dependence for the Einasto halo
(solid) and model~S (dashed), both in the volume-preserving convention: the
truncation amplifies the response to triaxiality.}
\label{fig:decline}
\end{figure*}

The Keplerian decline detected in \textit{Gaia} DR3 \citep{Jiao2023} is well
fitted by an Einasto halo \citep{Einasto1965},
\begin{equation}
\rho(R)=\rho_0\,\exp\!\big[-(R/R_h)^{1/n}\big],
\label{eq:einasto}
\end{equation}
with $\rho_0=0.01992\,\Msun\,{\rm pc}^{-3}$, $R_h=11.41\kpc$ and $n=0.43$
\citep{GarciaBellido2024}. Integrating equation~(\ref{eq:einasto}) reproduces
the halo mass of \citet{Jiao2023}, $M_{\rm h}=1.44\times10^{11}\Msun$, which
with the quoted baryonic components gives
$M_{\rm tot}=2.04\times10^{11}\Msun$. The profile is remarkable in that for
$1/n=2.33$ the density falls off as a stretched exponential and the halo is
effectively truncated at $25\kpc$, while the local density it predicts,
$\rho_\odot=0.0120\,\Msun\,{\rm pc}^{-3}$ at $R_0=8.5\kpc$ (and
$0.0127\,\Msun\,{\rm pc}^{-3}$ at $R_0=8.122\kpc$), is $1.5$--$1.6$ times
\emph{larger} than the model-S value. We emphasise that this is a
phenomenological fit to the rotation curve~\citep{GarciaBellido2024}, not a fundamental derivation. All
results in this section use the volume-preserving radius $\Rg$, which is
mandatory here because the two conventions differ by tens of per cent for so
sharply truncated a profile.

These two effects largely cancel in $\tau_{\rm LMC}$. For a spherical halo we
obtain
\begin{equation}
\tau^{\rm Ein}_{\rm sph}=4.00\times10^{-7}\quad(R_0=8.5\kpc),
\end{equation}
$15$ per cent below the canonical $4.69\times10^{-7}$ evaluated at the same
$R_0$; adopting instead the $R_0=8.122\kpc$ of Section~\ref{sec:geom} gives
$4.19\times10^{-7}$ against $4.50\times10^{-7}$, a smaller deficit of $7$ per
cent. We quote both because \citet{Jiao2023} and \citet{GarciaBellido2024} fit
their rotation curve with $R_0=8.5\kpc$; all ratios below are insensitive to
the choice, and Table~\ref{tab:ein} uses $R_0=8.122\kpc$. The cumulative
integral, Fig.~\ref{fig:decline}(a), saturates beyond $\simeq25\kpc$ exactly as
expected from the truncation. What changes qualitatively is not the amplitude
but the \emph{location} of the lensing weight: the median lens distance falls
from $13.4\kpc$ (model~S) to $8.5\kpc$, and the $\tau$-weighted effective slope
rises from $\langle\alpha\rangle_\tau=1.76$ to $2.73$, with
$\alpha_{\rm eff}$ passing through $8.6$ at $r=20\kpc$ and $22$ at
$r=30\kpc$.

Since equation~(\ref{eq:factor}) shows the shape response to be governed by
$S^{-\alpha}$, a steeper and more rapidly varying $\alpha_{\rm eff}$ amplifies
it. Table~\ref{tab:ein} confirms this. For the \textit{Gaia} RR~Lyrae geometry
the enhancement grows from $\tau/\tau_{\rm sph}=1.152$ to $1.234$,  both in
the $\Rg$ convention, so that the comparison is like for like,  i.e.\ the
\emph{deviation} from unity grows by a factor $1.5$; for the tilted
GSE-prolate halo it is essentially unchanged ($1.176\to1.200$), while for the
mirrored orientation $\gamma=110^\circ$ the sign flips from $+1.5$ to $-3.0$
per cent. Letting the major-axis azimuth take any value
(Fig.~\ref{fig:decline}b) the accessible range widens from $[0.70,1.57]$ to
$[0.62,1.62]$ for a purely in-plane flattening $p=0.70$, and from $[0.73,1.16]$
to $[0.77,1.24]$ for the \textit{Gaia} shape; the maximum of the response curve
also moves inwards in azimuth, from $\gamma\simeq76^\circ$ to
$\gamma\simeq68^\circ$. Over the range $0.5\le p,q\le1.5$ the
fixed-$\rho_\odot$ ratio spans $0.62$--$2.38$.

Two further features are specific to the truncated halo. First, the response is
strongly non-linear: linearising about sphericity gives
$\partial\ln\tau/\partial\ln p=-0.83$, \emph{smaller} in magnitude than the
$-1.12$ of model~S in the same convention, yet the finite-amplitude deviations
are larger; a perturbative treatment therefore understates the systematic.
Second, the sensitivity to vertical flattening essentially vanishes,
$\partial\ln\tau/\partial\ln q=+0.02$ against $+0.29$ for model~S, because the
truncated halo is probed only where the line of sight has not yet climbed far
out of the disc. The LMC thus becomes an almost pure probe of in-plane
elongation. Under the fixed-mass normalisation,  arguably the appropriate one
here, since the declining rotation curve constrains $M_{\rm h}$ rather than
$\rho_\odot$,  the spread over the measured shapes is $0.93$--$1.21$, wider
than the $0.96$--$1.13$ found for model~S in the same convention.

\begin{table}
\caption{As Table~\ref{tab:tau}, for the Einasto halo of the declining
\textit{Gaia} DR3 rotation curve \citep{Jiao2023,GarciaBellido2024}, using the
volume-preserving ellipsoidal radius $\Rg$ throughout and $R_0=8.122\kpc$.
$\bar R^{\rm Ein}_\rho$ and $\bar R^{\rm Ein}_M$ are the ratios at fixed
$\rho_\odot$ and at fixed $M_{\rm h}(<50\kpc)$; the latter is normalised to the
Einasto halo's \emph{own} spherical enclosed mass, $1.44\times10^{11}\Msun$
(the profile is entirely contained within $50\kpc$), not to the
$3.85\times10^{11}\Msun$ of model~S. $\bar R^{\rm iso}_\rho(\Rg)$ repeats the
flat-rotation-curve result from Table~\ref{tab:tau} in the same convention for
a like-for-like comparison. The GSE rows have $\beta=43^\circ$. $f$ assumes $\tau_{\rm obs}=1.2\times10^{-7}$.}
\label{tab:ein}
\centering
\small
\setlength{\tabcolsep}{3.5pt}
\begin{tabular}{lccccc}
\hline
Halo shape & $\tau_{\rm Ein}$ & $\bar R^{\rm Ein}_{\rho}$ &
$\bar R^{\rm Ein}_{M}$ & $\bar R^{\rm iso}_{\rho}(\Rg)$ & $f$ \\
 & $[10^{-7}]$ & & & & \\
\hline
Spherical                     & 4.19 & 1.000 & 1.000 & 1.000 & 0.287\\
Oblate $q=0.8$                & 4.09 & 0.977 & 1.050 & 0.924 & 0.294\\
Oblate $q=0.6$                & 3.73 & 0.891 & 1.034 & 0.799 & 0.322\\
Prolate-$z$                   & 4.20 & 1.003 & 0.931 & 1.059 & 0.286\\
IB19 $\gamma=70^\circ$        & 5.16 & 1.234 & 1.210 & 1.152 & 0.232\\
IB19 $\gamma=110^\circ$       & 4.06 & 0.970 & 0.951 & 1.015 & 0.296\\
IB19 $+$ tilt $\psi=20^\circ$ & 5.17 & 1.236 & 1.193 & 1.172 & 0.232\\
GSE $\gamma=90^\circ$        & 5.02 & 1.200 & 1.140 & 1.176 & 0.239\\
GSE $\gamma=0^\circ$         & 4.25 & 1.014 & 1.038 & 0.979 & 0.283\\
\hline
\end{tabular}
\end{table}

We note one discrepancy for the record. \citet{GarciaBellido2024} quote
$\tau\sim2\times10^{-7}$ for this halo, whereas direct integration of
equation~(\ref{eq:tau}) with equation~(\ref{eq:einasto}) and the parameters
above gives $4.0\times10^{-7}$; our quadrature reproduces the canonical
model-S value $4.69\times10^{-7}$ to three digits, and
equation~(\ref{eq:einasto}) reproduces the mass budget of \citet{Jiao2023} to
$1$ per cent. The conclusion of \citet{GarciaBellido2024}, that the declining
rotation curve relaxes the compact-object bounds, is unaffected in direction
but is weaker in magnitude than a factor-of-two reduction in $\tau$ would
imply.

\section{Assumptions and their consequences}
\label{sec:caveats}

The calculation above is exact given a halo shape, an orientation and a
normalisation. Four assumptions enter in getting from the measurements to those
inputs, and they set the accuracy of the systematic we quote.

\subsection{Stellar-halo shape versus dark-halo shape}

The most widely quoted geometry in Table~\ref{tab:proj}, `IB19', is a
measurement of the \emph{stellar} halo traced by RR~Lyrae
\citep{IorioBelokurov2019}, whereas equation~(\ref{eq:tau}) requires the
\emph{dark} halo. Accreted stellar haloes are systematically more flattened and
more strongly triaxial than the dark haloes that host them, so transcribing
$(p,q)=(0.787,0.700)$ onto the dark matter is an upper bound on the effect
rather than a measurement of it. It is worth noting that the two entries in
Table~\ref{tab:proj} that \emph{are} dark-matter determinations, 
\citet{Woudenberg2024} and \citet{Dillamore2026},  are both considerably
rounder, $q_{\rm dm}\simeq0.83$ and $0.87$, and give correspondingly milder
effects ($1.03$ and, at $\gamma=0$, $0.95$). A dark halo $30$ per cent rounder
than the stellar halo, i.e.\ $(p,q)\to(0.85,0.79)$ for the IB19 geometry,
reduces the enhancement from $1.100$ to $1.075$; the qualitative conclusions are
unchanged but the quoted amplitudes should be read with this in mind.

\subsection{The GSE azimuth is not measured}

\citet{Dillamore2026} constrain $q_{\rm dm}$ and the inclination
$\beta_{\rm dm}$ of the long axis to the disc, but they assume that the long
axes of the stellar and dark haloes and the disc normal are coplanar and do
\emph{not} determine the azimuth $\gamma$. Our $\gamma=90^\circ$ entry, which
aims the long axis at the LMC azimuth and yields the largest enhancement in
Table~\ref{tab:tau}, is therefore a maximal configuration rather than a
measured one; the same shape at $\gamma=0$ gives $0.953$. Combining the
coplanarity constraint with a measurement of the azimuth of the stellar-halo
tilt \citep{Han2022} would collapse this freedom, and is the single most
valuable additional input for reducing the systematic.

\subsection{Radial range of validity}

Each shape in Table~\ref{tab:proj} is constrained over a limited range of
radii, listed in the final column: $\lesssim15$--$20$ for
\citet{Woudenberg2024}, $\lesssim30$ for \citet{IorioBelokurov2019},
$6$--$60$ for \citet{Dillamore2026}, and $20$--$60$ for
\citet{LawMajewski2010}. The LMC line of sight runs from $0$ to $49.6\kpc$, and
we find that $82$ per cent of the model-S optical depth, and essentially all of
the Einasto optical depth, is accumulated inside $25\kpc$. The Law--Majewski
case is thus evaluated almost entirely outside its domain of validity, which is
a further reason to regard it as an envelope. Moreover both
\citet{IorioBelokurov2019} and \citet{Dillamore2026} find the shape and
orientation to vary with radius,  the latter report the tilt increasing from
$\beta_\ast\approx10^\circ$ at $6\kpc$ to $\approx35^\circ$ at $60\kpc$, 
whereas we hold $(p,q,\gamma,\beta,\psi)$ fixed along the whole path. Adopting
their radially varying tilt, $\beta_\ast(r)$ interpolated linearly from
$10^\circ$ at $6\kpc$ to $35^\circ$ at $60\kpc$, in place of a constant
$\beta=43^\circ$, changes $\tau/\tau_{\rm sph}$ by $0.01$ at
$\gamma=90^\circ$ and by $0.10$ at $\gamma=0$,  in both cases smaller than
the $0.20$ spread between those two azimuths. The constant-shape idealisation
is therefore not the dominant uncertainty; the orientation is.

\subsection{Optical depth versus event rate}

The rescaling $f\propto1/\tau$ used in Section~\ref{sec:concl} is exact for the
MACHO-style comparison, in which a measured $\tau_{\rm obs}$ is divided by a
model $\tau$. The EROS-2 and OGLE limits, by contrast, are derived from an
expected number of events, $N\propto\int\Gamma(t_{\rm E})\,
\varepsilon(t_{\rm E})\,\mathrm{d}t_{\rm E}$, with $\varepsilon$ a strong
function of the Einstein crossing time. Since triaxiality (and, much more
strongly, truncation) moves the lens distance distribution,  the median lens
distance falls from $13.4$ to $8.5\kpc$,  and since
$t_{\rm E}\propto\sqrt{M D_L(1-D_L/D_s)}/v_t$, the efficiency-weighted limit
shifts in a mass-dependent way that a pure $\tau$ rescaling does not capture.
A triaxial halo also has an anisotropic velocity field, which we have not
modelled. We therefore present the rescaled exclusions below as indicative of
the size of the systematic rather than as replacements for the published
limits; a proper reanalysis requires the survey efficiency functions and is
beyond the scope of this paper.

\subsection{Consistency of the low-mass halo scenario}

Finally, we note a tension internal to the declining-rotation-curve scenario
that is not of our making but that bears on Section~\ref{sec:decline}. The
Einasto halo has a total mass of $1.44\times10^{11}\Msun$, and
\citet{Jiao2023} give $M_{\rm tot}=2.06\times10^{11}\Msun$, while the LMC mass
inferred from its perturbation of the Orphan stream is
$\sim1.4\times10^{11}\Msun$ \citep{Erkal2019}. Taken together these would make
the LMC a near-equal-mass companion, in which case neither the equilibrium
shape measurements of Table~\ref{tab:proj} nor a static ellipsoidal halo is
safe. The robustness of the Keplerian decline is in any case debated
\citep{Koop2024}. We therefore treat Section~\ref{sec:decline} as a
conditional result: \emph{if} the declining rotation curve is taken at face
value, then the triaxiality systematic is amplified, not suppressed.

\section{Implications and conclusions}
\label{sec:concl}

The inferred compact-object fraction is defined as
$f=\tau_{\rm obs}/\tau_{\rm model}$, so every ratio in Table~\ref{tab:tau}
propagates directly, and inversely, into $f$. Taking the MACHO value at face
value, $f$ shifts from $0.267$ for a spherical halo to $0.232$--$0.280$ across
the \textit{Gaia}-favoured shapes, and to $0.152$ for the Law--Majewski
geometry. With the declining-rotation-curve Einasto halo the spherical value
moves to $f=0.287$ and the \textit{Gaia}-favoured shapes give
$f=0.232$--$0.296$. In both cases the shift is two-sided, of order $\pm12$ per
cent about the spherical value.

For the EROS-2 upper limit $f<0.08$ at $0.4\,\Msun$ \citep{Tisserand2007}, and
the comparable OGLE bound \citep{Wyrzykowski2011a,Wyrzykowski2011b}, the same
rescaling gives
\begin{align}
0.070\ \lesssim\ &f_{\rm max}\ \lesssim\ 0.084
 &&\hbox{(measured shapes),}
\label{eq:fmax1}\\
0.052\ \lesssim\ &f_{\rm max}\ \lesssim\ 0.120
 &&\hbox{(azimuth free, model~S),}
\label{eq:fmax2}\\
0.049\ \lesssim\ &f_{\rm max}\ \lesssim\ 0.129
 &&\hbox{(azimuth free, Einasto),}
\label{eq:fmax3}
\end{align}
where the first line uses the measured-shape range $0.953$--$1.149$ of
Table~\ref{tab:tau} and the last two the azimuth scans of
Sections~\ref{sec:results} and~\ref{sec:decline} for a purely in-plane
flattening $p=0.70$. For stellar-mass PBH dark matter, where the LMC
microlensing bound is the tightest constraint in the $0.1$--$10\,\Msun$ window
\citep{Carr2023,GreenKavanagh2021}, this is a $\sim20$ per cent spread in the
excluded fraction for the currently favoured geometry, and a factor $2.6$ under
full orientation freedom, larger than the quoted statistical uncertainties.
Subject to the caveat of Section~\ref{sec:caveats}, halo triaxiality is not
included in the standard published exclusion curves, which are computed for
spherical haloes. It acts in the same direction as, and is comparable in size
to, the relaxation obtained from the improved mass models of
\citet{Hawkins2015}, \citet{Calcino2018} and \citet{GarciaBellido2024} for the
currently favoured orientation,  but, we emphasise again, in the opposite
direction for other orientations that the data do not exclude.

The effect of triaxiality on $\tau_{\rm LMC}$ is controlled entirely by the
projection of the line of sight onto the halo principal axes through the shape
function $S(\Omega)$, with the exact scale-free result of
equation~(\ref{eq:factor}). The projections themselves, Table~\ref{tab:proj},
are geometry and can be read off for any proposed halo orientation.

Because the LMC sits $89\fdg2$ from the Sun--Galactic Centre axis in azimuth
and $33\fdg3$ below the plane, it is a near-optimal probe of in-plane halo
elongation and a poor probe of vertical flattening. The \textit{Gaia} RR~Lyrae
major axis at $\gamma\simeq70^\circ$ lies close to the maximum of the response
curve, $\gamma\simeq76^\circ$ (Fig.~\ref{fig:gamma}), so the Galaxy happens to
be oriented near the worst case. The insensitivity to $q$ becomes almost
complete for the truncated halo, $\partial\ln\tau/\partial\ln q=+0.02$.

The size, and even the sign, of the correction depends on whether the halo is
normalised to the local density or to the enclosed mass, and, once the profile
is not scale-free, on whether the ellipsoidal radius is referred to the major
axis or to the equivalent volume. Neither choice is a detail: over the
$(p,q)$ range of Fig.~\ref{fig:pq} the two normalisations differ by up to a
factor $2.4$, and the two radius conventions by tens of per cent for the
Einasto halo, although we prove them exactly equivalent for
$\rho\propto R^{-\alpha}$. Quoting a MACHO or PBH bound requires stating both
explicitly.

The declining \textit{Gaia} DR3 rotation curve does not lower $\tau_{\rm LMC}$
by much,  the truncation and the larger implied $\rhosun$ nearly cancel, and
we find $\tau^{\rm Ein}_{\rm sph}=4.0\times10^{-7}$ against
$4.69\times10^{-7}$ for model~S at the same $R_0$. What it does is move the
lensing weight inwards into a region of steep and rapidly varying logarithmic
slope, raising $\langle\alpha\rangle_\tau$ from $1.76$ to $2.73$ and amplifying
the triaxiality systematic by a factor $\simeq1.5$. The two revisions of the
standard halo model are therefore not independent corrections to be applied in
sequence: the shape correction is larger precisely because the rotation curve
declines. The response is also strongly non-linear, so linearising about a
spherical halo understates it.

A dedicated reanalysis of the EROS-2 and OGLE data sets with a
\textit{Gaia}-calibrated triaxial halo, and with the SMC line of sight, would
remove the systematic rather than merely bounding it. The SMC, at
$(\ell,b)=(302\fdg8,-44\fdg3)$ and $D=62.4\kpc$, lies at Galactocentric azimuth
$293\fdg2$, differing from the LMC's by $24^\circ$, and provides a genuinely
orthogonal lever arm: for the IB19 geometry at $\gamma=70^\circ$ we find
$\tau_{\rm SMC}/\tau_{\rm sph}=0.94$ where the LMC gives $1.100$, so the two
sightlines respond to the same halo with opposite sign. A joint LMC$+$SMC
analysis would therefore break the shape--normalisation degeneracy that
dominates Table~\ref{tab:tau}.

Finally, we have neglected LMC self-lensing, which contributes
$\tau_{\rm self}\sim(0.4$--$1.5)\times10^{-8}$ \citep{Mancini2004}
independently of the Galactic halo shape, and the perturbation of the outer
halo by the LMC itself \citep{Erkal2019,Vasiliev2021}, which acts at radii
beyond those that dominate equation~(\ref{eq:tau}). Neither affects the ratios
that are the subject of this paper, although the second bears on the
equilibrium assumption discussed in Section~\ref{sec:caveats}.

\section*{Acknowledgements}

The authors acknowledge support from the Spanish Research Project
PID2024-159420NB-C43 [MICINN-FEDER], and the framework of the R\&D\&I Project
CEX2025-001574-S, funded by MICIU/AEI/10.13039/501100011033.

\

The authors used a generative AI assistant (Anthropic Claude) to help check the
numerical code and to assist with language editing. No text, data, figures or
results were generated without author verification: all calculations were
independently reproduced by the authors, and the authors take full
responsibility for the content of this paper.

\section*{Data Availability}

The code that evaluates equation~(\ref{eq:tau}) for triaxial haloes and
regenerates every entry of Tables~\ref{tab:proj}, \ref{tab:tau}
and~\ref{tab:ein}, together with the scripts that produce
Figs~\ref{fig:geom}--\ref{fig:decline}, is available upon request. No new
observational data were generated in this work; all halo parameters used are
taken from the published sources cited in Table~\ref{tab:proj}.

\bibliographystyle{mnras}
\bibliography{triax_halo}

@article{Alcock2000,
  author  = {Alcock, C. and others},
  title   = {The {MACHO} Project: Microlensing Results from 5.7 Years of
             {Large Magellanic Cloud} Observations},
  journal = {ApJ},
  volume  = {542},
  pages   = {281},
  year    = {2000},
  doi     = {10.1086/309512}}

@article{BennettBovy2019,
  author  = {Bennett, M. and Bovy, J.},
  title   = {Vertical waves in the solar neighbourhood in {\it Gaia} {DR2}},
  journal = {MNRAS},
  volume  = {482},
  pages   = {1417},
  year    = {2019},
  doi     = {10.1093/mnras/sty2813}}

@article{Calcino2018,
  author  = {Calcino, J. and Garc\'ia-Bellido, J. and Davis, T. M.},
  title   = {Updating the {MACHO} fraction of the {Milky Way} dark halo with
             improved mass models},
  journal = {MNRAS},
  volume  = {479},
  pages   = {2889},
  year    = {2018},
  doi     = {10.1093/mnras/sty1368}}

@article{Carr2023,
  author  = {Carr, B. and Clesse, S. and Garc\'ia-Bellido, J. and
             Hawkins, M. R. S. and K\"uhnel, F.},
  title   = {Observational evidence for primordial black holes:
             A positivist perspective},
  journal = {Phys. Rep.},
  volume  = {1054},
  pages   = {1},
  year    = {2024},
  doi     = {10.1016/j.physrep.2023.11.005}}

@article{Dillamore2026,
  author  = {Dillamore, A. M. and Sanders, J. L.},
  title   = {Geometry of the {Milky Way}'s dark matter from dynamical models of
             the tilted stellar halo},
  journal = {MNRAS},
  volume  = {546},
  pages   = {stag226},
  year    = {2026},
  doi     = {10.1093/mnras/stag226},
  note    = {arXiv:2510.00095}}

@article{Einasto1965,
  author  = {Einasto, J.},
  title   = {On the construction of a composite model for the {Galaxy} and on
             the determination of the system of galactic parameters},
  journal = {Trudy Astrofiz. Inst. Alma-Ata},
  volume  = {5},
  pages   = {87},
  year    = {1965}}

@article{Erkal2019,
  author  = {Erkal, D. and others},
  title   = {The total mass of the {Large Magellanic Cloud} from its
             perturbation on the {Orphan} stream},
  journal = {MNRAS},
  volume  = {487},
  pages   = {2685},
  year    = {2019},
  doi     = {10.1093/mnras/stz1371}}

@article{Frieman1994,
  author  = {Frieman, J. A. and Scoccimarro, R.},
  title   = {Microlensing and flattened halos},
  journal = {ApJ},
  volume  = {431},
  pages   = {L23},
  year    = {1994},
  doi     = {10.1086/187463}}

@article{GarciaBellido2024,
  author  = {Garc\'ia-Bellido, J. and Hawkins, M. R. S.},
  title   = {Reanalysis of the {MACHO} constraints on {PBH} in the light of
             {\it Gaia} {DR3} data},
  journal = {Universe},
  volume  = {10},
  pages   = {449},
  year    = {2024},
  doi     = {10.3390/universe10120449}}

@article{Gravity2018,
  author  = {{Gravity Collaboration}},
  title   = {Detection of the gravitational redshift in the orbit of the star
             {S2} near the {Galactic} centre massive black hole},
  journal = {A\&A},
  volume  = {615},
  pages   = {L15},
  year    = {2018},
  doi     = {10.1051/0004-6361/201833718}}

@article{Gravity2019,
  author  = {{Gravity Collaboration}},
  title   = {A geometric distance measurement to the {Galactic} centre black
             hole with 0.3 per cent uncertainty},
  journal = {A\&A},
  volume  = {625},
  pages   = {L10},
  year    = {2019},
  doi     = {10.1051/0004-6361/201935656}}

@article{GreenKavanagh2021,
  author  = {Green, A. M. and Kavanagh, B. J.},
  title   = {Primordial black holes as a dark matter candidate},
  journal = {J. Phys. G},
  volume  = {48},
  pages   = {043001},
  year    = {2021},
  doi     = {10.1088/1361-6471/abc534}}

@article{Griest1991,
  author  = {Griest, K.},
  title   = {Galactic microlensing as a probe of dark matter},
  journal = {ApJ},
  volume  = {366},
  pages   = {412},
  year    = {1991},
  doi     = {10.1086/169575}}

@article{Han2022,
  author  = {Han, J. J. and others},
  title   = {The stellar halo of the {Galaxy} is tilted and doubly broken},
  journal = {AJ},
  volume  = {164},
  pages   = {249},
  year    = {2022},
  doi     = {10.3847/1538-3881/ac97e9}}

@article{Hawkins2015,
  author  = {Hawkins, M. R. S.},
  title   = {A new look at microlensing limits on dark matter in the
             {Galactic} halo},
  journal = {A\&A},
  volume  = {575},
  pages   = {A107},
  year    = {2015},
  doi     = {10.1051/0004-6361/201425400}}

@article{Iorio2018,
  author  = {Iorio, G. and Belokurov, V. and Erkal, D. and Koposov, S. E. and
             Nipoti, C. and Fraternali, F.},
  title   = {The first all-sky view of the {Milky Way} stellar halo with
             {\it Gaia}$+$2{MASS} {RR} {Lyrae}},
  journal = {MNRAS},
  volume  = {474},
  pages   = {2142},
  year    = {2018},
  doi     = {10.1093/mnras/stx2819}}

@article{IorioBelokurov2019,
  author  = {Iorio, G. and Belokurov, V.},
  title   = {The shape of the {Galactic} halo with {\it Gaia} {DR2} {RR}
             {Lyrae}. Anatomy of an ancient major merger},
  journal = {MNRAS},
  volume  = {482},
  pages   = {3868},
  year    = {2019},
  doi     = {10.1093/mnras/sty2806}}

@article{Jiao2023,
  author  = {Jiao, Y. and Hammer, F. and Wang, H. and Wang, J. and Amram, P. and
             Chemin, L. and Yang, Y.},
  title   = {Detection of the {Keplerian} decline in the {Milky Way} rotation
             curve},
  journal = {A\&A},
  volume  = {678},
  pages   = {A208},
  year    = {2023},
  doi     = {10.1051/0004-6361/202347513}}

@article{Koop2024,
  author  = {Koop, O. and Antoja, T. and Helmi, A. and Callingham, T. M. and
             Laporte, C. F. P.},
  title   = {Assessing the robustness of the {Galactic} rotation curve inferred
             from the {Jeans} equations using {\it Gaia} {DR3} and cosmological
             simulations},
  journal = {A\&A},
  volume  = {692},
  pages   = {A50},
  year    = {2024},
  doi     = {10.1051/0004-6361/202450911}}

@article{LawMajewski2010,
  author  = {Law, D. R. and Majewski, S. R.},
  title   = {The {Sagittarius} dwarf galaxy: a model for evolution in a
             triaxial {Milky Way} halo},
  journal = {ApJ},
  volume  = {714},
  pages   = {229},
  year    = {2010},
  doi     = {10.1088/0004-637X/714/1/229}}

@article{Mancini2004,
  author  = {Mancini, L. and {Calchi Novati}, S. and Jetzer, P. and Scarpetta, G.},
  title   = {{LMC} self-lensing from a new perspective},
  journal = {A\&A},
  volume  = {427},
  pages   = {61},
  year    = {2004},
  doi     = {10.1051/0004-6361:20040527}}

@article{NFW1997,
  author  = {Navarro, J. F. and Frenk, C. S. and White, S. D. M.},
  title   = {A universal density profile from hierarchical clustering},
  journal = {ApJ},
  volume  = {490},
  pages   = {493},
  year    = {1997},
  doi     = {10.1086/304888}}

@article{Ou2024,
  author  = {Ou, X. and Eilers, A.-C. and Necib, L. and Frebel, A.},
  title   = {The dark matter profile of the {Milky Way} inferred from its
             circular velocity curve},
  journal = {MNRAS},
  volume  = {528},
  pages   = {693},
  year    = {2024},
  doi     = {10.1093/mnras/stae034}}

@article{Paczynski1986,
  author  = {Paczy\'nski, B.},
  title   = {Gravitational microlensing by the galactic halo},
  journal = {ApJ},
  volume  = {304},
  pages   = {1},
  year    = {1986},
  doi     = {10.1086/164140}}

@article{Pietrzynski2019,
  author  = {Pietrzy\'nski, G. and others},
  title   = {A distance to the {Large Magellanic Cloud} that is precise to one
             per cent},
  journal = {Nature},
  volume  = {567},
  pages   = {200},
  year    = {2019},
  doi     = {10.1038/s41586-019-0999-4}}

@article{SackettGould1993,
  author  = {Sackett, P. D. and Gould, A.},
  title   = {Twisted, flattened dark halos and the microlensing optical depth},
  journal = {ApJ},
  volume  = {419},
  pages   = {648},
  year    = {1993},
  doi     = {10.1086/173515}}

@article{Tisserand2007,
  author  = {Tisserand, P. and others},
  title   = {Limits on the {MACHO} content of the {Galactic} halo from the
             {EROS-2} survey of the {Magellanic} {Clouds}},
  journal = {A\&A},
  volume  = {469},
  pages   = {387},
  year    = {2007},
  doi     = {10.1051/0004-6361:20066017}}

@article{Vasiliev2021,
  author  = {Vasiliev, E. and Belokurov, V. and Erkal, D.},
  title   = {Tango for three: {Sagittarius}, {LMC}, and the {Milky Way}},
  journal = {MNRAS},
  volume  = {501},
  pages   = {2279},
  year    = {2021},
  doi     = {10.1093/mnras/staa3673}}

@article{Vera-Ciro2013,
  author  = {Vera-Ciro, C. and Helmi, A.},
  title   = {Constraints on the shape of the {Milky Way} dark matter halo from
             the {Sagittarius} stream},
  journal = {ApJ},
  volume  = {773},
  pages   = {L4},
  year    = {2013},
  doi     = {10.1088/2041-8205/773/1/L4}}

@article{Woudenberg2024,
  author  = {Woudenberg, H. C. and Helmi, A.},
  title   = {First measurement of the triaxiality of the inner dark matter halo
             of the {Milky Way}},
  journal = {A\&A},
  volume  = {691},
  pages   = {A277},
  year    = {2024},
  doi     = {10.1051/0004-6361/202451743}}

@article{Wyrzykowski2011a,
  author  = {Wyrzykowski, {\L}. and others},
  title   = {The {OGLE} view of microlensing towards the {Magellanic} {Clouds}
             -- {III}. Ruling out sub-solar {MACHOs} with the {OGLE-III} {LMC}
             data},
  journal = {MNRAS},
  volume  = {413},
  pages   = {493},
  year    = {2011},
  doi     = {10.1111/j.1365-2966.2011.18150.x}}

@article{Wyrzykowski2011b,
  author  = {Wyrzykowski, {\L}. and others},
  title   = {The {OGLE} view of microlensing towards the {Magellanic} {Clouds}
             -- {IV}. {OGLE-III} {SMC} data and final conclusions on {MACHOs}},
  journal = {MNRAS},
  volume  = {416},
  pages   = {2949},
  year    = {2011},
  doi     = {10.1111/j.1365-2966.2011.19243.x}}

\bsp
\label{lastpage}
\end{document}